\documentclass[10pt,aps,pre,twocolumn,notitlepage,superscriptaddress,preprintnumbers]{revtex4-1}
\usepackage{amsmath,amssymb,amsfonts}
\usepackage{bm}
\usepackage{graphicx,color}
\usepackage{verbatim}
\usepackage{float}
\usepackage[caption = false]{subfig}
\usepackage{dcolumn}
\usepackage{soul}
\usepackage{natbib}
\usepackage{hyperref}
\usepackage{cleveref}
\usepackage{appendix}
\usepackage[utf8x]{inputenc}

\begin{document}
	\title{Hybrid balance theory: Heider balance under higher order interactions}
	
	\author{M. H. Hakimi Siboni}
	\email{m.hosein.h.s@gmail.com}
	\affiliation{Department of Physics, Shahid Beheshti University, Evin, Tehran $ 1983969411 $, Iran}
	
	\author{A. Kargaran}
	\email{amir.kargaran25@gmail.com}
	\affiliation{Department of Physics, Shahid Beheshti University, Evin, Tehran $ 1983969411 $, Iran}
	
	\author{G. R. Jafari}
	\email{g_jafari@sbu.ac.ir}
	\affiliation{Department of Physics, Shahid Beheshti University, Evin, Tehran $ 1983969411 $, Iran}
	\affiliation{Irkutsk State Technical University, Lermontova Str. $ 83 $, $ 664074 $ Irkutsk, Russia}

	
	\begin{abstract}
		Heider's Balance Theory in signed networks, which consists of friendship or enmity relationships, is a model that relates the type of relationship between two people to the third person. In this model, there is an assumption of the independence of triadic relations, which means that the balance or imbalance of one triangle does not affect another and the energy only depends on the number of each type of triangle. There is evidence that in real network data, in addition to third-order interactions (Heider Balance), higher-order interactions also play a role. One step beyond the Heider Balance, the effect of Quartic Balance has been studied by removing the assumption of triangular independence. Application of quartic balance results in the influence of the balanced or imbalanced state of neighboring triangles on each specific one. Here, a question arises that how the Heider Balance is affected by the existence of Quartic Balance (fourth-order). The phase diagram obtained from the mean-field method shows there is a threshold for higher-order interaction strength, below which a third-order interaction dominates and there are no imbalance triangles in the network, and above this threshold, squares effectively determine the balance state in which the imbalance triangles can survive. The solution of the mean-field indicates that we have a first-order phase transition in terms of random behavior of agents (temperature) which is in accordance with the Monte Carlo simulation results.
	\end{abstract}
	
	\maketitle
	\section{Introduction}	
	In 1946 Fritz Heider, an Austrian psychologist, introduced Heider Balance Theory (HBT), a social theory which studies relations among society by considering triadic interactions between agents of the society \cite{heider1}. In HBT, Heider suggests that society evolves in a manner to reduce psychological stress among the triadic relations \cite{heider2}. Later on, Cartwright and Harary \cite{cartwright}, using the concepts of graph theory, expanded the balance theory and introduced the concept of structural balance, since then HBT has been applied to many diverse fields of study including sociology \cite{szell,altafini,saeedian,sheykhali, hassanibesheli, oloomi,mahsa,masoumi, malarz, belaza1, belaza2, kirkely}, ecology \cite{saiz}, cognitive sciences \cite{zahra, majid}, biology \cite{abbas, saberi} and international relations \cite{hart}\cite{galam}. In HBT, the relationship between two people is affected by the information that passes from one intermediary person, and the question that arises is why only one person should be considered as an intermediary. In fact, the complexity of social networks may require us to consider more intermediaries in our modeling. In order to bring up the idea of several intermediaries, we need to consider higher-order interactions alongside triadic Heider Balance.
	
	In HBT a society is modeled by a graph in which every node represents a member of the society and the link between two nodes shows the relation between them. In this model, links representing friendship have a value of $ +1 $ and links which show enmity have a value of $ -1 $. Every triad in the graph according to the sign of its constituent links can be balanced or imbalanced. Triads with three positive links $ [+++] $ or two negative links and a positive one $ [--+] $ (the enemy of your friend is your enemy) are balanced triads and triads with three negative links $ [---] $, or a negative link and two positive ones $ [++-] $ (your friends are enemies of each other) are imbalanced. This means that if the product of the sign of the links in a triad is positive (negative) that triad is balanced (imbalanced). The imbalanced triads of a society tend to become balanced, and the whole society will reach the structural balance when all links become positive (heaven) or the society becomes bipolar. By bipolar we mean a society containing two rival groups in which all relations inside each group are friendship, and all the relations between the groups are enmity.
	
	Using a diversity of methods people have studied different aspects of HBT, including Antal \textit{et al.} \cite{antal1}\cite{antal2} who studied (discrete-time) dynamics of HBT and its transitions, by introducing and studying two kinds of dynamics based on Heider Balance. In their work, they considered two discrete values for links ($ \pm 1 $). Scientists with a similar method have examined the phase transition and the emergence of multipolar societies \cite{montakhab1}\cite{montakhab2}. In another work By considering the values of the links to be a real number, which not only represents the type of relations but also the strength of them, Ku\l akowski \textit{et al.} introduced a differential equation for the dynamics of the graph based on HBT and studied dynamics of HBT more detailedly \cite{kulakowski}. Later Marvel \textit{et al.}, using the proposed differential equation of Ku\l akowski studied the structural balance of a network more deeply \cite{marvel2}. In another work, Marvel \textit{et al.} introduced a Hamiltonian for networks governed by social balance and studied their energy landscape and local minima\cite{marvel1}. Introducing the Hamiltonian was important progress in the field and it allowed people to use equilibrium statistical physics approaches to study signed networks based on HBT. Using this approach and by introducing a concept of temperature Rabbani \textit{et al.}, worked on the effects of temperature on HBT and studied the phase transitions of this system \cite{fereshteh}.
	
	HBT has been applied to real-world data and in some cases, people have shown that HBT is an acceptable tool for analyzing some sort of data \cite{facchetti}, but in some other researches people claimed that a large number of real-world networks are not structurally balanced and HBT can not explain them, so it seems that HBT is not yet a complete and reliable theory to apply on real-world data \cite{leskovec,estrada2,estrada3}. Many different approaches are proposed to fulfill HBT drawbacks. Physicists generalized the Heider balance model by considering a dynamic role for individuals (nodes),  for example, G\'orski \textit{et al.} proposing $G$ attribute for each node and presenting a model that unites the basic principles of the HBT and homophily \cite{janusz}. They examined the final states of their model for a different number of attributes ($G$). Recently, the equilibrium statistical mechanics approach for this interaction on networks with different node attributes has been investigated \cite{pham}. Scientists studied special cases of this model in which people have only two or three types of beliefs with different methods \cite{thurner, singh, amir2}. Another way to generalize this model is to consider higher-order interactions. Kargaran \textit{et. al.} \cite{amir1} by presenting the Quartic Balance Theory considered fourth-order interactions in which squares, instead of triangles, play an influential role. They showed in addition to two structural balances states (heaven and bipolar) this model has two other states of hell and negative bipolar and also studied the effect of temperature on the network.
	
\begin{figure}
		\centering
		\includegraphics[width=0.8\linewidth]{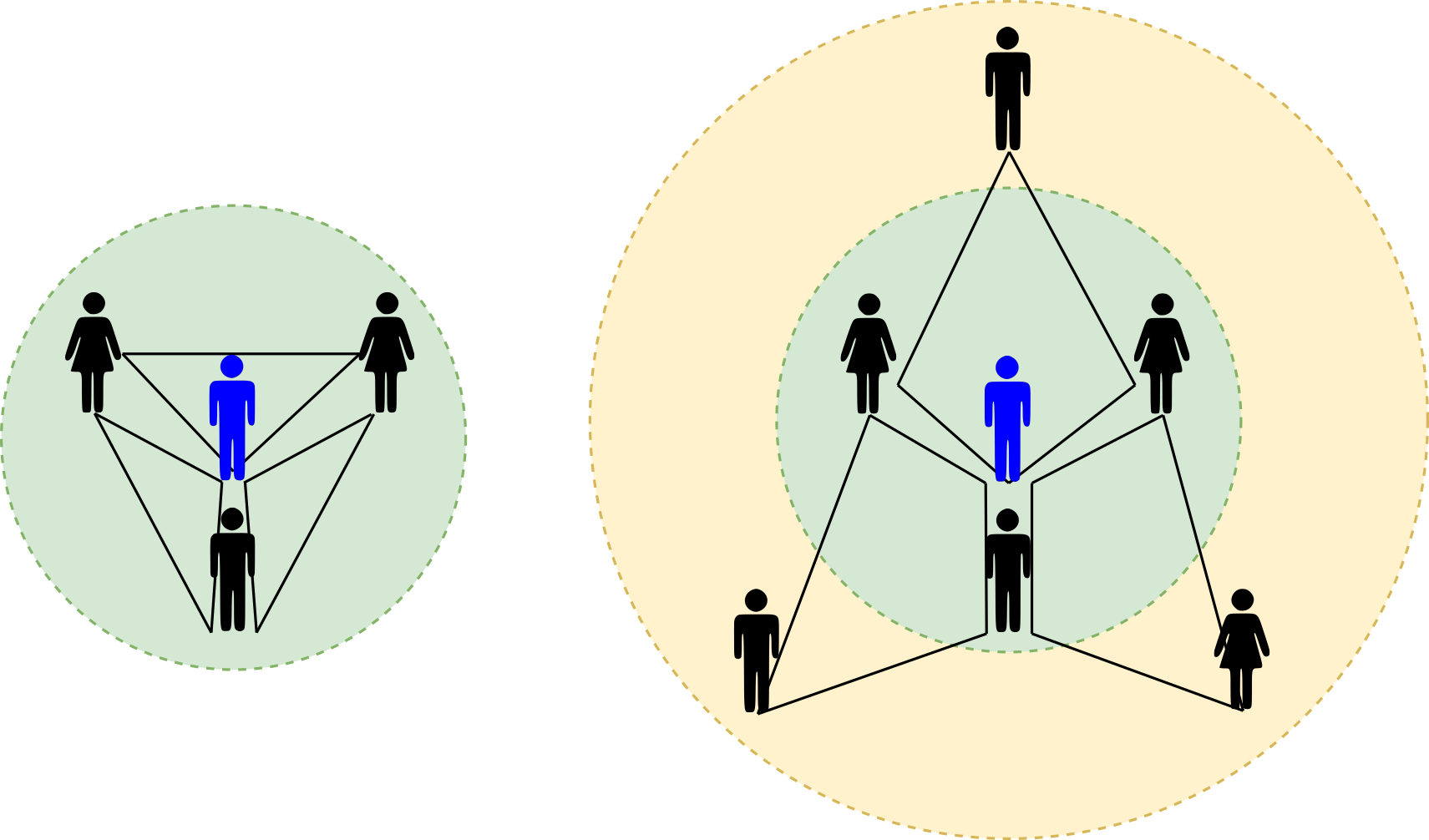}
		\caption{Left: Triadic interactions between an agent (blue) and its neighbors. Right: Quartic interactions between an agent and other agents that do not have direct interactions necessarily. In our model, both interactions exist.}
		\label{fig:fig1}
	\end{figure}

	In this work, we want to propose a model in which both triadic and quartic interactions play a role (Fig.~\ref{fig:fig1}). That means the social tension in the network is a result of tensions of both triadic and quartic interactions, each with a specified strength, and the system tends to minimize this mixed tension instead of reducing triadic or quartic tensions separately. We also propose a concept of temperature for this system and using the concepts of equilibrium statistical physics, \textit{exponential random graphs} (ERG) \cite{cimini} and the mean-field theory tries to study this system and find its stable states, phase transitions, and its dependence on the strength of quartic relations in the mixed tension. We also confirm our analytical results with the outcomes of Monte Carlo simulations.
	
	\section{Model}
	As discussed above here we consider a society in which Heider and quartic balance play a role \cite{fereshteh}\cite{amir1}. The overall social tension of this model is considered as
	
	\begin{equation}\label{Hamiltonian}
	\begin{aligned}
	\mathcal{H}(G)&=-t(G)-g\; s(G)\\
	&=-\sum_{i<j<k}\sigma_{ij}\sigma_{jk}\sigma_{ki}-g\sum_{i<j<k<\ell}\sigma_{jk}\sigma_{ki}\sigma_{j\ell}\sigma_{\ell i}.
	\end{aligned}
	\end{equation}
	
	In physics, these types of cost functions are known as \textit{Hamiltonian} which assigns a social tension (or energy) to each network configuration $ (G) $. In Eq.~\ref{Hamiltonian}, ${\sigma_{ij}}$ is an element of the symmetric adjacency matrix $(\sigma_{ij}=\sigma_{ji})$, that can be $+1$ (friendship) or $-1$ (enmity). The first and second terms of Eq.~\ref{Hamiltonian} each have their minimum energy configurations. The first term (Heider balance), which was introduced in \cite{marvel1} has two balanced states called heaven and bipolar(I) (known as bipolar in the literature). In the heaven state, all social ties are friendly. In the bipolar(I) state, two poles are formed whose internal relations are friendship and the relations between the poles are enmity. The second term (quartic balance) has four balanced states (minimum energy), two of them are like the Heider balance minimums (heaven and bipolar). The other two are all-enemy (hell) and bipolar(II) states. In bipolar(II) state, all the links within poles are enmity, and the links between them are friendly. The parameter $g$ is a real number between $0$ and $1$ which determines the strength of quartic terms against triadic ones and we call it combination coefficient from now on.
	
	We used the concept of social temperature $(T)$ in our model. This parameter allows social relations to be changed randomly with Boltzmann probability. These random changes may reduce or increase the energy or overall social tension and prevent the system from getting stuck in the local minima of overall social tension (jammed states).   
	
	To study this problem analytically we make use of the \textit{exponential random graphs} \cite{holland, strauss1, snijders1, robins, cranmer, snijders2, newman1, newman2, newman3}. As a consequence the probability of finding a network at the specific configuration $G$ can be written as $ P(G) \propto e^{-\beta \mathcal{H}(G)} $ where $ \beta = 1/T $. In the following sections by using this probability and the method of the mean-field we are going to study fixed points of the system, behavior of order parameters of it, and its phase transition. We finally compare all the results with computer simulations.
	
	\section{Analysis}	
	\subsection{Mean-field Approximation}
	In this section we apply mean-field approximation to find average quantities. First quantity is mean of links $ \langle\sigma_{ik}\rangle $ .We extract all the terms in our Hamiltonian that involve $ \sigma_{jk} $ like
	\begin{equation}\label{local-feild}
	-\mathcal{H}_{jk}=\sigma_{jk}\left[\sum_{\mu\neq j,k}\sigma_{j\mu}\sigma_{\mu k}+g\sum_{\mu\neq j,k}\sum_{\nu\neq j,k}\sigma_{j\mu}\sigma_{\mu\nu}\sigma_{\nu k}\right].
	\end{equation}
	We also call all the terms in Hamiltonian that don't involve $ \sigma_{jk} $, by $ \mathcal{H'} $, so the Hamiltonian is $ \mathcal{H} =\mathcal{H'}+\mathcal{H}_{kj}$. By analogy with spin models we can call Eq.~\ref{local-feild} the \textit{local field} that is coupled to spin $ \sigma_{jk} $. For calculating the mean value we should start with
	\begin{equation}
	\langle\sigma_{jk}\rangle=\sum_{G}\sigma_{jk}\mathcal{P}(G),
	\end{equation}  
	where $ P(G) = e^{-\beta\mathcal{H}(G)} /\mathcal{Z} $ is probability of the configuration $ G $ and $ \mathcal{Z} $ is the partition function, and we have
	\begin{equation}\label{single}
	\begin{aligned}
	&\langle\sigma_{jk}\rangle=\frac{1}{\mathcal{Z}}\sum_{\{\sigma\neq\sigma_{jk}\}}e^{-\beta\mathcal{H}'}\sum_{\sigma_{jk}=\pm 1}\sigma_{jk}e^{-\beta\mathcal{H}_{jk}} \\
	&=\frac{\sum_{\{\sigma\neq\sigma_{jk}\}}e^{-\beta\mathcal{H}'}\left[e^{-\beta\mathcal{H}_{jk}(\sigma_{jk}=+1)}-e^{-\beta\mathcal{H}_{jk}(\sigma_{jk}=-1)}\right]}{\sum_{\{\sigma\neq\sigma_{jk}\}}e^{-\beta\mathcal{H}'}\left[e^{-\beta\mathcal{H}_{jk}(\sigma_{jk}=+1)}+e^{-\beta\mathcal{H}_{jk}(\sigma_{jk}=-1)}\right]}\\
	&=\frac{\left\langle e^{-\beta\mathcal{H}_{jk}(\sigma_{jk}=+1)}-e^{-\beta\mathcal{H}_{jk}(\sigma_{jk}=-1)}\right\rangle_{G'}}{\left\langle e^{-\beta\mathcal{H}_{jk}(\sigma_{jk}=+1)}+e^{-\beta\mathcal{H}_{jk}(\sigma_{jk}=-1)}\right\rangle_{G'}},
	\end{aligned}
	\end{equation}
	where $ \langle\cdots\rangle_{G'} $ is average on all configurations that not have $ \sigma_{jk} $. Now we use mean-field approximation to replace spin variables in Eq.~\ref{single} with their ensemble averages like $ q\equiv\langle \sigma_{jk}\sigma_{ki}\rangle $ and $ o\equiv\langle \sigma_{ki}\sigma_{jl}\sigma_{li}\rangle $, which is  the mean value of two stars and open squares. By two star we mean two links that have a common node and  and by open square we mean a square with one missing link. By defining $ p\equiv \langle\sigma_{jk}\rangle$ we can  write
	\begin{equation}\label{meanofedges}
	p=\tanh\Big(\beta(n-2)\,q+\beta\,g(n-2)(n-3)\,o\Big),
	\end{equation}
	where $ n $ is the number of nodes of our network. The coefficients depending on $ n $ in Eq.~\ref{meanofedges} are related to the number of triangles and squares which share $ \sigma_{jk} $ in the fully connected network. The number of triangles that have a common link is equal to $ n-2 $ and also the number of squares that have a shared link is equal to $ 2\times\binom{n-2}{2} $. The constant two  is related to the two possible configurations for a square which share $ \sigma_{jk} $ with fixed nodes. Now we want to find the mean of two stars ($ q $). First we should find all terms that $ \sigma_{jk}\sigma_{ki} $. For finding this purpose we should extract all terms that contain $\sigma_{jk} $ and $ \sigma_{ki} $ plus all terms that involve $ \sigma_{jk}\sigma_{ki} $ which is
	
	\begin{widetext}
		\begin{equation}
		\begin{aligned}
		-\mathcal{H}_{\vee}&=\sigma_{jk}\left[\sum_{\mu\neq i,j,k}\sigma_{j\mu}\sigma_{\mu k}+g\sum_{\mu,\nu\neq i,j,k}\sigma_{j\mu}\sigma_{\mu\nu}\sigma_{\nu k}\right]
		+\sigma_{ki}\left[\sum_{\mu\neq i,j,k}\sigma_{k\mu}\sigma_{\mu i}+g\sum_{\mu,\nu\neq i,j,k}\sigma_{k\mu}\sigma_{\mu\nu}\sigma_{\nu i}\right]\\
		&\qquad\qquad\qquad\qquad\qquad\qquad\qquad\qquad\qquad\qquad\qquad\		\qquad+\sigma_{ij}\sigma_{jk}\sigma_{ki}+g\sigma_{jk}\sigma_{ki}\sum_{\mu\neq i,j,k}\sigma_{i\mu}\sigma_{\mu j}.
		\end{aligned}
		\end{equation}
		We can find  $ \langle\sigma_{jk}\sigma_{ki}\rangle $ by calculating $ \sum_{G}\sigma_{jk}\sigma_{ki}\mathcal{P}(G) $ 
		using the method that we mentioned, we have
		
		\begin{equation}
		\begin{aligned}
		\langle\sigma_{jk}\sigma_{ki}\rangle&=\frac{1}{\mathcal{Z}}\sum_{\{\sigma\}}\sigma_{jk}\sigma_{ki}e^{-\beta\mathcal{H}(G)}=\frac{\sum_{\{\sigma\neq\sigma_{jk},\sigma_{ki}\}}e^{-\beta\mathcal{H}'}\sum_{\{\sigma_{jk},\sigma_{ki}=\pm 1\}}\sigma_{jk}\sigma_{ki}e^{-\beta\mathcal{H}_{\vee}}}{\sum_{\{\sigma\neq\sigma_{jk},\sigma_{ki}\}}e^{-\beta\mathcal{H}'}\sum_{\{\sigma_{jk},\sigma_{ki}=\pm 1\}}e^{-\mathcal{H}_{\vee}}}\\
		&=\frac{\left\langle e^{-\beta\mathcal{H}_{\vee}(\sigma_{jk}=1,\sigma_{ki}=1)}-e^{-\beta\mathcal{H}_{\vee}(\sigma_{jk}=1,\sigma_{ki}=-1)}-e^{-\beta\mathcal{H}_{\vee}(\sigma_{jk}=-1,\sigma_{ki}=1)}+e^{-\beta\mathcal{H}_{\vee}(\sigma_{jk}=-1,\sigma_{ki}=-1)}\right\rangle_{G'}}{\left\langle e^{-\beta\mathcal{H}_{\vee}(\sigma_{jk}=1,\sigma_{ki}=1)}+e^{-\beta\mathcal{H}_{\vee}(\sigma_{jk}=1,\sigma_{ki}=-1)}+e^{-\beta\mathcal{H}_{\vee}(\sigma_{jk}=-1,\sigma_{ki}=1)}+e^{-\beta\mathcal{H}_{\vee}(\sigma_{jk}=-1,\sigma_{ki}=-1)}\right\rangle_{G'}}.
		\end{aligned}
		\end{equation}	
		The $ \langle\cdots\rangle_{G'} $ means that the average is taken on all configurations that do not contain $ \sigma_{jk}$ and $\sigma_{ki}$. The results of mean-field approximation is
		\begin{equation}\label{meanoftwostars}
		q=\frac{e^{2\beta(n-3)q+2\beta g(n-3)(n-4)+\beta p +\beta g(n-3)q}-2 e^{-\beta p-\beta g (n-3) q}+e^{-2\beta(n-3)q-2\beta g(n-3)(n-4)+\beta p +\beta g(n-3)q}}{e^{2\beta(n-3)q+2\beta g(n-3)(n-4)+\beta p +\beta g(n-3)q}+2 e^{-\beta p-\beta g (n-3) q}+e^{-2\beta(n-3)q-2\beta g(n-3)(n-4)+\beta p +\beta g(n-3)q}}.
		\end{equation}
		We calculated the mean of open squares, triangles, squares, and the energy with the same analogy in Appendix \ref{appendix:mean quantity}.
	\end{widetext}
	By substituting Eq.~\ref{meanofedges} in  Eq.~\ref{meanoftwostars} and Eq.~\ref{opensquares} we can find the coupled self-consistence equations as
	\begin{equation}\label{self-consistence}
	\begin{aligned}
	q&= f(q,\,o\,;\;\beta,\,g,\,n),\\
	o&= h(q,\,o\,;\;\beta,\,g,\,n).
	\end{aligned}
	\end{equation}
	\begin{figure}[t]
		\centering
		\includegraphics[scale=0.2]{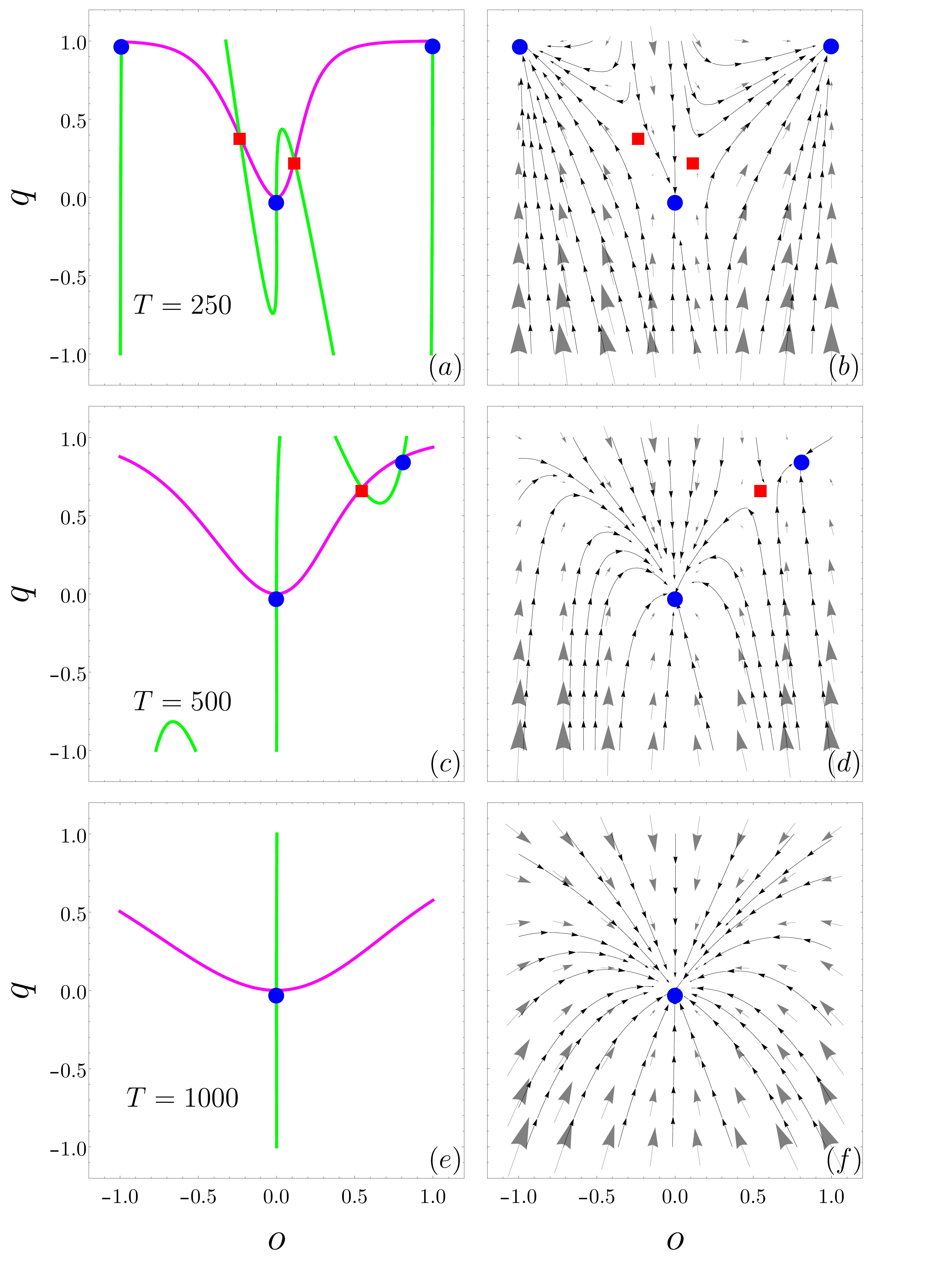}
		\caption{(Color online) Left column: simultaneous solution of Eq.~\ref{self-consistence} for $ g=0.1 $ and $ T=250, 500, 1000 $. Right column: the flow diagram for visualizing the convergence (divergence) from stable (unstable) fixed points. The red square and blue circle indicate the stable and unstable fixed points respectively. The number of nodes is $ n=100 $.}
		\label{fig:fig2}
	\end{figure}
	\begin{figure}
		\centering
		\includegraphics[scale=0.35]{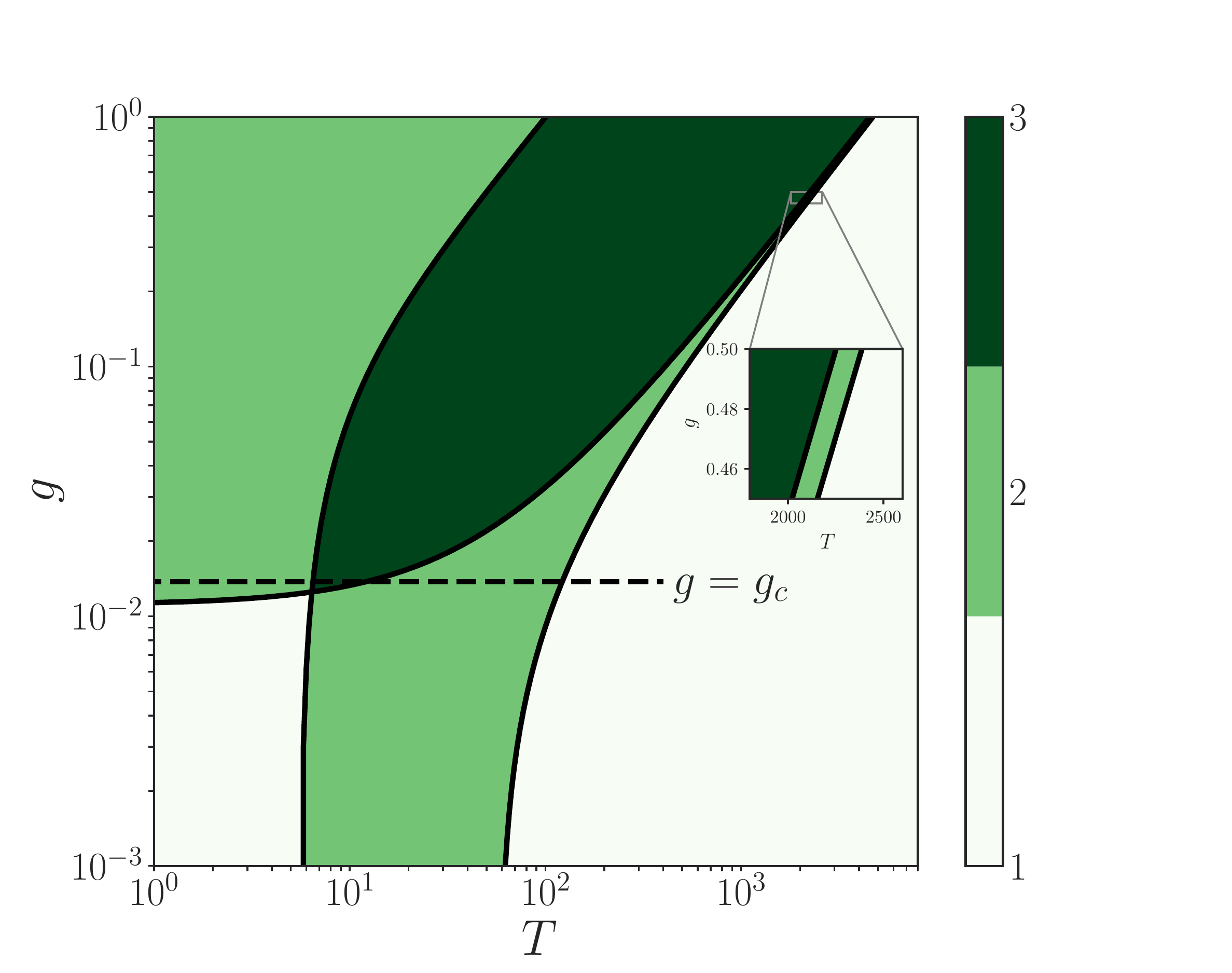}
		\caption{The phase diagram for for the number of simultaneous solutions for Eq.~\ref{self-consistence} in $ (T, g) $ space. This diagram divides into two regions by $ g_c $. The region $ g>g_c $ the squares have more energy than triangles and have a dominant role and in region $ g<g_c $ the triangles have the dominant role. Inset: the two-solution region becomes very narrow in log-login the log-log plot but it exists in the linear plot with approximately fixed width. The number of nodes is $ n=100 $.} 
		\label{fig:fig3}
	\end{figure}
	
	We are looking for simultaneous solutions of Eq.~\ref{self-consistence}. In the Fig.~\ref{fig:fig1} (a), (c) and (e) we have drawn the solutions of above equations in all allowed intervals ($-1\le q\le 1$, $-1\le o\le 1$) with specified $ g=0.1 $ and temperature. The intersections of these lines are the simultaneous solutions or fixed points of Eq.~\ref{self-consistence} (disks and squares). We find that in low temperatures we have five solutions that turn to three solutions for higher temperatures $ T<T_c $, below the critical temperature $ T_c $ at $ T<T_c $. Above this temperature, at $ T>T_c $ we have just one solution. The number of solutions depends on our model's free parameters which are $ g $ and $ T $.
	
	In the Fig.~\ref{fig:fig1} (b), (d) and (f) we have determined the stability and instability of the simultaneous solutions of Eq.~\ref{self-consistence}. We assume a point very close to the fixed points $ (q^*+\delta q,o^*+\delta o) $ that this point is mapped by equations Eq.~\ref{self-consistence} to a new point like
	\begin{equation}\label{linear}
	\begin{aligned}
	q^*+\delta q'&= f(q^*+\delta q,\,o^*+\delta o\,;\;\beta,\,g,\,n),\\
	o^*+\delta o'&= h(q^*+\delta q,\,o^*+\delta o\,;\;\beta,\,g,\,n).
	\end{aligned}
	\end{equation}
	For $ \delta q\ll 1 $ and $ \delta o\ll 1 $ the Taylor expansion for above equations are
	\begin{equation}
	\begin{aligned}
	q^*+\delta q'&\approx f(q^*,o^*\,;\;\beta,\,g,\,n)\,
	+\, \frac{\partial f}{\partial q}\Big |_{\substack{q=q^*\\o=o^*}}\delta q \, +\, \frac{\partial f}{\partial o}\Big |_{\substack{q=q^*\\o=o^*}}\delta o,\\
	o^*+\delta o'&\approx h(q^*,o^*\,;\;\beta,\,g,\,n)\,
	+\, \frac{\partial h}{\partial q}\Big |_{\substack{q=q^*\\o=o^*}}\delta q \, +\, \frac{\partial h}{\partial o}\Big |_{\substack{q=q^*\\o=o^*}}\delta o.
	\end{aligned}
	\end{equation}
	The linearised transformation of above equations is
	\begin{equation}\label{matrixeq}
	\left(\begin{array}{c} \delta q'\\ \delta o' \end{array}\right)  = \mathbf{J}\left(\begin{array}{c} \delta q \\ \delta o \end{array}\right), 
	\end{equation}
	where $\mathbf{J}$ is the Jacobian matrix
	\begin{equation}
	\mathbf{J}=
	\left(\begin{array}{cc} \partial f/\partial q & \partial f/\partial o\\ \partial h/\partial q & \partial h/\partial o \end{array}\right)_{\substack{q=q^*\\o=o^*}}.
	\end{equation}
	By finding the diagonalized form of above equation which is
	\begin{equation}\label{digonal-form}
	\left(\begin{array}{c} \delta_{d}q' \\ \delta_{d}o' \end{array}\right)  =
	\left(\begin{array}{cc}\lambda_1 & 0\\ 0 & \lambda_2 \end{array}\right)\left(\begin{array}{c} \delta_{d} q \\ \delta_{d} o \end{array}\right),
	\end{equation}
	we can find the stability condition of the fixed points. If the magnitude of both eigenvalues are bigger than one ($ |\lambda_1|>1 $ and $ |\lambda_2|>1 $) the fixed point is unstable and if $ |\lambda_1|<1 $ and $ |\lambda_2|<1 $, the fixed point is stable. Using this method, it can be understood that in stable fixed points (unstable fixed points) the new difference ($ \delta_{d}q' $ and $ \delta_{d}o' $) decreases (increases) with each iteration. For better understanding of the stability and instability of fixed points, the flow diagram of Eq.~\ref{self-consistence} are also illustrated in Fig.~\ref{fig:fig2} (b), (d) and (f). The flow converges to the stable fixed points and diverges from unstable ones. The components of this vector field (flow) are as follows
	\begin{equation}
	\begin{aligned}
	u&\equiv f(q,\,o\,;\;\beta,\,g,\,n)-q,\\
	v&\equiv h(q,\,o\,;\;\beta,\,g,\,n)-o.\\
	\end{aligned}
	\end{equation}
	
\begin{figure}[t]
		\centering
		\includegraphics[scale=0.85]{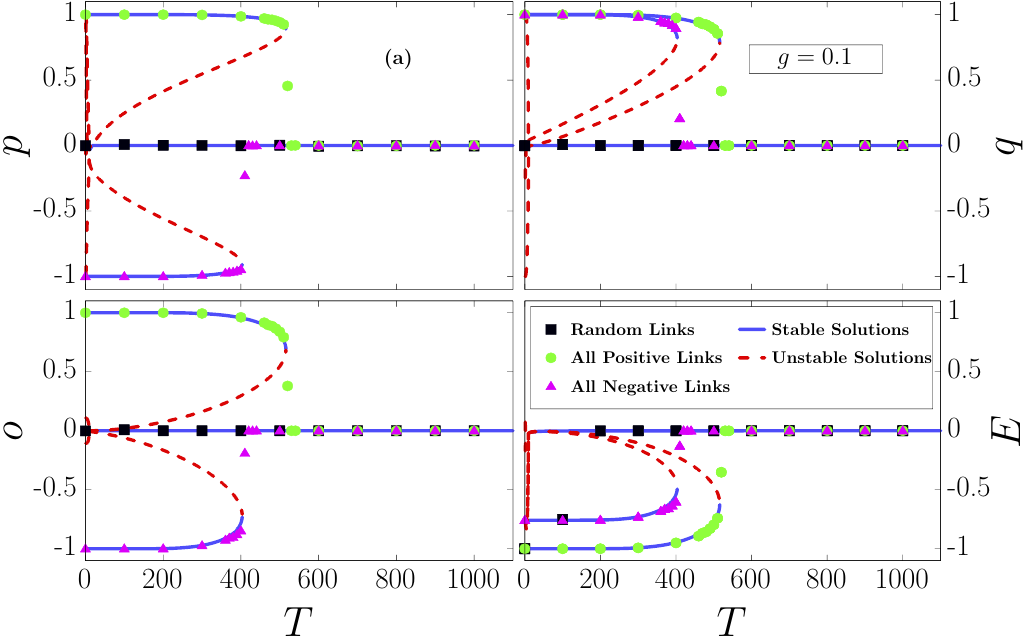}
		\includegraphics[scale=0.85]{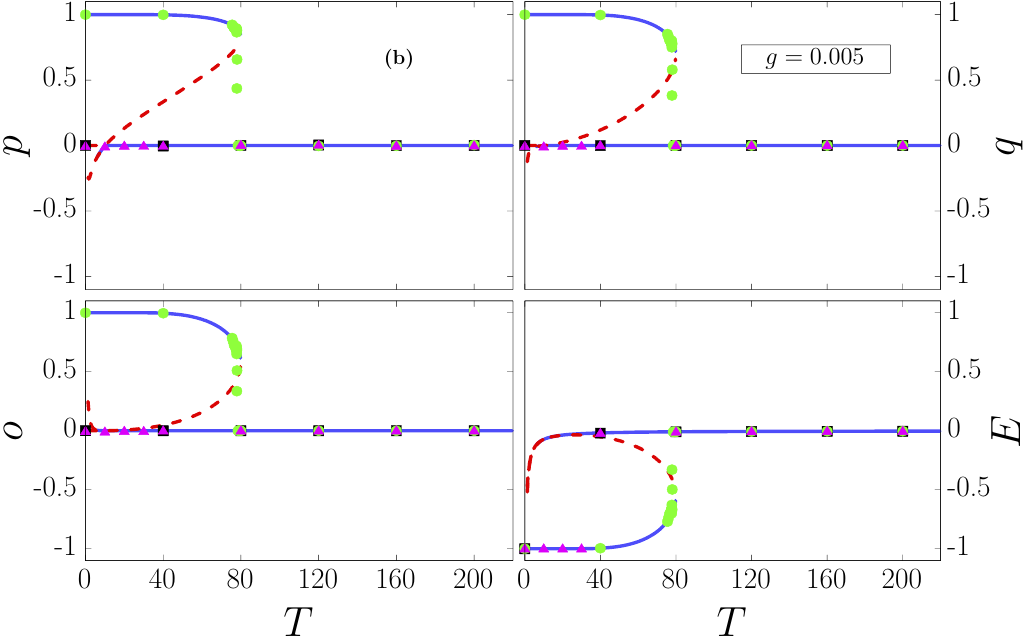}
		\caption{a: Simultaneous solutions of Eq.~\ref{self-consistence} and simulation results for g = 0.1 and $0 \leq T \leq 1100$. b: Simultaneous solutions of Eq.~\ref{self-consistence} and simulation results for g = 0.051 and $0 \leq T \leq 220$.}
		\label{fig:fig4}
	\end{figure}

	The stable fixed points have properties that can be discussed separately. According to Fig.~\ref{fig:fig4} (a) the trivial solution of Eq.~\ref{self-consistence}  is $ (q^{*}=0,\;o^{*}=0) $, which is a stable fixed point for a  long-range of temperatures. This fixed point describes the random phase of the network in which there is no order. Fixed points $ (q^*=+1,\, o^*=+1) $ is the minimum energy (balanced) state for triplet and quartic interaction (first and second term of the Hamiltonian) that is known as heaven in the literature of social physics. The last stable fixed point  $ (q^*=+1,\, o^*=-1) $ is stable only in quartic interaction which is introduced as hell \cite{amir1}. By changing the temperature (Fig.~\ref{fig:fig2} (c), (d)), the number and value of stable fixed points on the $ q-o$ plane change.
	
	In Fig.~\ref{fig:fig3} we plot the number of stable fixed points versus the free parameters of the model ($ T $ and $ g $). Given that the number of squares and triangles in the fully connected network is exactly known (number of squares is $ \binom{n}{4} $ and number of triangles is $ \binom{n}{3} $), we can find a certain value for $ g $ that the energy of all triangles and all squares of the network, are exactly equal. In this article, we call this specific value $ g_c $. For the values greater than $ g_c $, squares contribute more to the total energy than triangles and vice versa. In other words in $ g>g_c $ the squares of the network are dominant and in $ g<g_c $ the triangles are dominant \cite{fereshteh}\cite{amir1}. The value of the $ g_c $ can be calculated as a function of the network size as
	\begin{equation}
	g_c=\frac{\binom{n}{3}}{3\times\binom{n}{4}}=\frac{4}{3(n-3)}.
	\end{equation}
	The coefficient $ 3 $ came from three square configurations from four selected nodes. In Fig.~\ref{fig:fig2} we have shown this value by a horizontal dashed line for a network with $ n=100 $.
	
	\subsection{Simulations}
	To simulate our model we use a fully connected network with $n=100$ nodes. Every link in the network can take a value of $+1$ for friendly relations, or $-1$ for unfriendly relations. We performed the Monte Carlo simulations for fixed combination coefficients $(g)$ ($g_1=0.1>g_c$ and $g_2=0.005<g_c$) at different temperatures ($T$). The initial conditions used in this simulation are all links positive, all links negative and a random configuration, respectively. The update rules for each time step are:
	
	\begin{enumerate}
		\item Pick a link at random by choosing two random nodes $(i \neq j)$.
		\item Attempt a flip of the link, so if it is a friendship it would be changed to an enmity and vice versa.
		\item Calculate the change in the energy of the network $\Delta{E}$ using Hamiltonian Eq.~\ref{Hamiltonian}
		\item Accept or reject the flip according to the following rules
		\begin{enumerate}
			\item If $\Delta{E} < 0$ accept the flip.
			\item if $\Delta{E} \geq 0$ choose a random real value $r$ between $0$ and $1$ and if $r < P = e^{(-\beta{\Delta{E}})}$ accept the flip and reject it otherwise.
		\end{enumerate}
		\item If accepted, update the network configuration and the energy of the system.
		\item Continue to the next time step from step 1.
	\end{enumerate}
	These simulations are shown in Fig.~\ref{fig:fig4}.
	
	We compare our simulation and theory together in Fig.~\ref{fig:fig4} for two values of combination factor (g) which are above and below the $g_c$ ($g_1=0.1>g_c$ in Fig.~\ref{fig:fig4} (a) and $g_2=0.005<g_c$ Fig.~\ref{fig:fig4} (b)). Our analytical calculations using the mean-field approximation (Fig.~\ref{fig:fig4}) show that there are four thermal regions for combination factors bigger than $g_c$ ($g_1$)  and three below it  ($g_2$).
	
	The first region for both $g_1$ and $g_2$ is the cold region where thermal noise is not that much to randomize the system and the system is in perfect order. These states in a regime where squares dominate ($g>g_c$) are heaven ($o=1$, $q=1$) and hell ($o=-1$, $q=1$) however in a regime where triangles dominate ($g<g_c$), only heaven is the perfectly ordered state. The initial conditions of all positive (negative) edges can lead us to the ordered state of heaven (hell) and results are in good agreement with our theory (green circle and magenta triangle).
	
	In the second region ($10<T<400$ for $g_1$ and $6<T<80$ for $g_2$)  as thermal noise is increased, besides previous solutions, another stable solution can be seen for both systems. This state is completely random at high temperatures and ordered at low temperatures. In the social physics literature, this state is called bipolar at low temperatures. The mean quantities for this state are  zero ($q=0$, $o=0$). In the simulation, this state can be reached by random initial condition (black square in Fig.~\ref{fig:fig4}) and results fit well with analytical calculations for a long range of temperature. The reason that the mean-field approximation does not predict the bipolar state in low temperature is that in this state the homogeneity of the field applied to each link is lost, which is the basic assumption of this method.
	
	In the third thermal region for $g_1$ ($400<T<513$), our analytical calculations show that the critical temperature of the regular state hell ($o=-1$, $q=1$) is lower than heaven ($o=1$, $q=1$). In other words, the state of heaven shows more resistance to randomization. The reason for this is that when we are in the states of heaven or hell, the quartic interaction with both of these states is at its minimum energy, while the triplet interaction is only with the state of heaven at its minimum energy, so the triplet interaction for the state of hell acts like temperature effectively, and the order of this state is lost at a lower temperature than heaven. This feature causes that in the all-positive initial condition (that leads to the final state of heaven), the critical temperature is greater than the critical temperature of the state of hell (with the initial condition of all-negative links). Our simulations with the initial conditions of all positive or negative links are completely consistent with the theory. In the third thermal region for $g_2$ ($6<T<80$), our theory predicts two stable answers that are confirmed by different initial conditions in the simulation. Here, because we are in the triangle dominant regime, we have a temperature for critical temperature ($T_c$), which, unlike the square dominant regime, is not related to the initial conditions. 
	
	Finally, the fourth temperature region for $g_1$  ($513<T$) and the third temperature zone for $g_2$  ($80<T$) is the random zone. In this region, the temperature is so large that no order lasts and only the random state is the answer to the equations ($q=0$, $o=0$). The critical temperature predicted by theory fits well with simulation for both combination factors (g).
	
\begin{figure}
		\centering
		\includegraphics[width=0.9\linewidth]{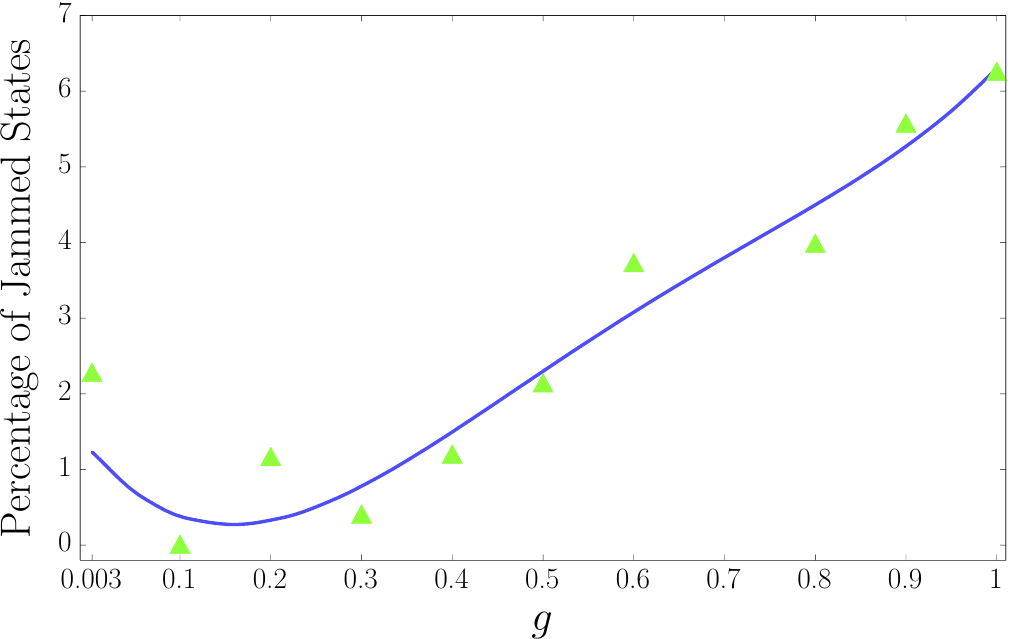}
		\caption{Percentage of jammed states for a thousand ensembles with different $ g $ values.}
		\label{fig:Fig6}
	\end{figure}

	As a final calculation, we studied jammed states in our model. For different values of $ g $ at zero temperature we simulated the system $ 10^3 $ times, counted the number of jammed states, and calculated their percentage. As can be seen in Fig.~\ref{fig:Fig6} at very low values of g (triadic regime) we have jammed states. As $ g $ increases the ratio of jammed states decreases until it reaches its minimum at $ g = 0.1 $. Afterward, as quartic interactions are in dominance, the ratio of jammed states increases monotonically.
	
	\section{Conclusions}
	
	In real data networks, the Heider Balance Theory is partially true \cite{abbas, zahra, majid, estrada2} and these networks have imbalanced triangles and higher-order interactions are ignored. With this in mind, one can ask whether the interaction of the triplets (triadic) gives us enough information and by considering higher-order interactions how the structural balance will change. In this work, we studied HBT in the presence of a fourth-order (quartic) interaction and learned that there is a threshold for higher-order interaction below which the HBT gives a correct idea of ​​the system so that there are no unbalanced triangles in structural balance. But above this threshold, the HBT is challenged and a new structural balance emerges which consists of unbalanced triangles. These significant differences between model HBT and the proposed model indicate the importance of higher-order interaction, which requires further investigations. 
	
	\begin{acknowledgments}                
		We want to thanks M. Ghanbarzadeh for comments that improved the manuscript.
	\end{acknowledgments}                
	\appendix
	\begin{appendices}
		\section{Mean Values of Open Squares}
		\label{appendix:mean quantity}
		In this section we calculate the average of open squares. As previous calculations, we extract all terms that contain parts of a special open square, and we write the Hamiltonian as $ \mathcal{H}=\mathcal{H}'+ \mathcal{H}_{\sqcup}$, where
		\small
		\begin{equation}
		\begin{aligned}
		-\mathcal{H}_{\sqcup}&=\sigma_{jk}\left[\sum_{\mu\neq i,j,k,\ell}\sigma_{j\mu}\sigma_{\mu k}+g\sum_{\mu,\nu\neq i,j,k,\ell}\sigma_{j\mu}\sigma_{\mu\nu}\sigma_{\nu k}\right]\\
		&+\sigma_{ki}\left[\sum_{\mu\neq i,j,k,\ell}\sigma_{k\mu}\sigma_{\mu i}+g\sum_{\mu,\nu\neq i,j,k,\ell}\sigma_{k\mu}\sigma_{\mu\nu}\sigma_{\nu i}\right]\\
		&+\sigma_{j\ell}\left[ \sum_{\mu\neq i,j,k,\ell}\sigma_{j\mu}\sigma_{\mu \ell}+g\sum_{\mu,\nu\neq i,j,k,\ell}\sigma_{j\mu}\sigma_{\mu\nu}\sigma_{\nu \ell}\right]\\
		&+2\sigma_{jk}\sigma_{ki}\sigma_{ij} +2\sigma_{jk}\sigma_{k\ell} \sigma_{\ell j}+ \sigma_{ki}\sigma_{i\ell}\sigma_{\ell k}+ \sigma_{ji}\sigma_{i\ell}\sigma_{\ell j}\\
		&+ g\sigma_{jk}\sigma_{ki}\sum_{\mu\neq i,j,k,\ell}\sigma_{i\mu}\sigma_{\mu j}
		+ g\sigma_{jk}\sigma_{j\ell}\sum_{\mu\neq i,j,k,\ell}\sigma_{k\mu}\sigma_{\mu\ell}\\
		&+g\left(\sigma_{jk}\sigma_{ki}\sigma_{j\ell}\right)\sigma_{\ell i}.\\
		\end{aligned}
		\end{equation}
		\normalsize
		The first three terms are all triangles and squares that have $ \sigma_{jk} $, $ \sigma_{ki} $, and $ \sigma_{j\ell} $ respectively. Subsequent terms appear from counting the number of triangles and squares correctly. For finding the average of open squares by statistical mechanics we should calculate  
		\begin{equation}
		o\equiv\langle\sigma_{jk}\sigma_{ki}\sigma_{j\ell}\rangle=\sum_{\{\sigma_{jk},\sigma_{ki},\sigma_{j\ell}=\pm 1\}}\sigma_{jk}\sigma_{ki}\sigma_{j\ell}\mathcal{P}(G).
		\end{equation}
		We considered all configurations of the open squares with two link signs ($ \pm 1 $) and find the mean-field approximation for all configurations then find the average quantity. By defining $\Gamma(o)\equiv(n-5)(n-4)\,o$ we can write: 
		\small
		\begin{equation}
		\begin{aligned}
		-\mathcal{H}^{+++}_{\sqcup}&=3[g\Gamma(o)+(n-4)q]+2[g(n-4)q + 2p]+2q+gp,\\
		-\mathcal{H}^{---}_{\sqcup}&=-3[g\Gamma(o)+(n-4)q]+ 2[g(n-4)q+2p]-2q-gp,\\
		-\mathcal{H}^{-++}_{\sqcup}&=g\Gamma(o)-2[g(n-4)q+2p]+(n-4)q+2q-gp,\\
		-\mathcal{H}^{+-+}_{\sqcup}&=g\Gamma(o) +(n-4)q -gp,\\
		-\mathcal{H}^{++-}_{\sqcup}&=g\Gamma(o) +(n-4)q -gp,\\
		-\mathcal{H}^{--+}_{\sqcup}&=-g\Gamma(o) -(n-4)q + gp,\\
		-\mathcal{H}^{-+-}_{\sqcup}&=-g\Gamma(o) -(n-4)q + gp,\\
		-\mathcal{H}^{+--}_{\sqcup}&=-g\Gamma(o) - 2q - (n-4)q - 2[2p + g(n-4)q] +gp,\\
		\\
		\end{aligned}
		\end{equation}
		\normalsize
		where  $ \mathcal{H}^{abc}_{\sqcup}\equiv\mathcal{H}_{\sqcup}(\sigma_{jk}=a,\,\sigma_{ki}=b,\,\sigma_{j\ell}=c) $. Finally we have
		\begin{equation}\label{opensquares}
		o=\frac{u(p,\,q,\,o\,;\;\beta,\,g,\,n)}{v(p,\,q,\,o\,;\;\beta,\,g,\,n)}
		\end{equation}
		where
		\small
		\begin{equation}
		\begin{aligned}
		u(p,\,q,\,o\,;\;\beta,\,g,\,n)&=e^{\beta\mathcal{H}^{+++}_{\sqcup}}-e^{\beta\mathcal{H}^{---}_{\sqcup}}-e^{\beta\mathcal{H}^{-++}_{\sqcup}}-e^{\beta\mathcal{H}^{+-+}_{\sqcup}}\\
		&-e^{\beta\mathcal{H}^{++-}_{\sqcup}}+e^{\beta\mathcal{H}^{--+}_{\sqcup}}+e^{\beta\mathcal{H}^{-+-}_{\sqcup}}+e^{\beta\mathcal{H}^{+--}_{\sqcup}},\\
		v(p,\,q,\,o\,;\;\beta,\,g,\,n)&=e^{\beta\mathcal{H}^{+++}_{\sqcup}}+e^{\beta\mathcal{H}^{---}_{\sqcup}}+e^{\beta\mathcal{H}^{-++}_{\sqcup}}+e^{\beta\mathcal{H}^{+-+}_{\sqcup}}\\
		&+e^{\beta\mathcal{H}^{++-}_{\sqcup}}+e^{\beta\mathcal{H}^{--+}_{\sqcup}}+e^{\beta\mathcal{H}^{-+-}_{\sqcup}}+e^{\beta\mathcal{H}^{+--}_{\sqcup}},
		\end{aligned}
		\end{equation}
		\normalsize
		and the mean of links ($ p $) calculated in Eq.~\ref{meanofedges}. 
		
		We can find the mean of triangles and squares with a similar method. In here we just write the required Hamiltonian for calculating these quantities. We have
		\small
		\begin{equation}
		\begin{aligned}
		-\mathcal{H}_{\Delta}&=\sigma_{ij}\left[ \sum_{\mu\neq i,j,k}\sigma_{i\mu}\sigma_{\mu j}+g\sum_{\mu,\nu\neq i,j,k}\sigma_{i\mu}\sigma_{\mu\nu}\sigma_{\nu j}\right]\\
		&+\sigma_{jk}\left[ \sum_{\mu\neq i,j,k}\sigma_{j\mu}\sigma_{\mu k}+g\sum_{\mu,\nu\neq i,j,k}\sigma_{j\mu}\sigma_{\mu\nu}\sigma_{\nu k}\right]\\
		&+\sigma_{ki}\left[ \sum_{\mu\neq i,j,k}\sigma_{k\mu}\sigma_{\mu i}+g\sum_{\mu,\nu\neq i,j,k}\sigma_{k\mu}\sigma_{\mu\nu}\sigma_{\nu i}\right]\\
		&+g\sigma_{jk}\sigma_{ki}\sum_{\mu\neq i,j,k}\sigma_{i\mu}\sigma_{\mu j}+g\sigma_{ij}\sigma_{jk}\sum_{\mu\neq i,j,k}\sigma_{k\mu}\sigma_{\mu i}\\
		&+g\sigma_{ji}\sigma_{ik}\sum_{\mu\neq i,j,k}\sigma_{k\mu}\sigma_{\mu j}+ \sigma_{ij}\sigma_{jk}\sigma_{ki},
		\end{aligned}
		\end{equation}
		\normalsize
		for triangles and squares we can write
		\small
		\begin{equation}
		\begin{aligned}
		-\mathcal{H}_{_{\square}}&=\sigma_{jk}\left[ \sum_{\mu\neq i,j,k,\ell}\sigma_{j\mu}\sigma_{\mu k}+g\sum_{\mu,\nu\neq i,j,k,\ell}\sigma_{j\mu}\sigma_{\mu\nu}\sigma_{\nu k}\right]\\
		&+\sigma_{ki}\left[ \sum_{\mu\neq i,j,k,\ell}\sigma_{k\mu}\sigma_{\mu i}+g\sum_{\mu,\nu\neq i,j,k,\ell}\sigma_{k\mu}\sigma_{\mu\nu}\sigma_{\nu i}\right]\\
		&+\sigma_{\ell i}\left[ \sum_{\mu\neq i,j,k,\ell}\sigma_{\ell\mu}\sigma_{\mu i}+g\sum_{\mu,\nu\neq i,j,k,\ell}\sigma_{\ell\mu}\sigma_{\mu\nu}\sigma_{\nu i}\right]\\
		&+\sigma_{j\ell}\left[ \sum_{\mu\neq i,j,k,\ell}\sigma_{j\mu}\sigma_{\mu \ell}+g\sum_{\mu,\nu\neq i,j,k,\ell}\sigma_{j\mu}\sigma_{\mu\nu}\sigma_{\nu \ell}\right]\\
		&+ \sigma_{jk}[\sigma_{ki}\sigma_{ij}+\sigma_{k \ell}\sigma_{\ell j}]+ \sigma_{ki}[\sigma_{\ell i}\sigma_{\ell k}+\sigma_{ij}\sigma_{jk}]
		\\
		&+ \sigma_{\ell i}[\sigma_{\ell k}\sigma_{ki}+\sigma_{ij}\sigma_{j\ell}]+ \sigma_{\ell j}[\sigma_{\ell k}\sigma_{kj}+\sigma_{ji}\sigma_{\ell i}]\\
		&+g\sigma_{jk}\sigma_{ki}\sum_{\mu\neq i,j,k,\ell}\sigma_{j\mu}\sigma_{\mu i}+g\sigma_{ki}\sigma_{i\ell}\sum_{\mu\neq i,j,k,\ell}\sigma_{k\mu}\sigma_{\mu\ell}\\
		&+g\sigma_{j\ell}\sigma_{\ell i}\sum_{\mu\neq i,j,k,\ell}\sigma_{j\mu}\sigma_{\mu i}+g\sigma_{jk}\sigma_{j\ell}\sum_{\mu\neq i,j,k,\ell}\sigma_{k\mu}\sigma_{\mu \ell}\\
		&+g\sigma_{jk}\sigma_{ki}\sigma_{jl}\sigma_{li}.\\
		\end{aligned}
		\end{equation}
		\normalsize
	\end{appendices}	
	
\end{document}